\documentclass[aps,prl,twocolumn,showpacs,preprintnumbers,amsmath,amssymb,superscriptaddress]{revtex4}%

\usepackage{graphicx}%
\usepackage{dcolumn}
\usepackage{amsmath}
\usepackage{color}
\usepackage{multirow}

\begin{document}


\title{Coherency of the superconducting state: the muon spin rotation and ARPES studies of
(BiPb)$_2$(SrLa)$_2$CuO$_{6+\delta}$}

\author{R.~Khasanov}
 \email[Corresponding author: ]{rustem.khasanov@psi.ch}
 \affiliation{Laboratory for Muon Spin Spectroscopy, Paul Scherrer
Institute, CH-5232 Villigen PSI, Switzerland}
\author{Takeshi~Kondo}
 \affiliation{Ames Laboratory and Department of Physics and Astronomy, Iowa State University, Ames, IA~50011, USA}
 \affiliation{Department of Crystalline Materials Science, Nagoya University, Nagoya 464-8603, Japan}
\author{M.~Bendele}
 \affiliation{Laboratory for Muon Spin Spectroscopy, Paul Scherrer
Institute, CH-5232 Villigen PSI, Switzerland}
 \affiliation{Physik-Institut der Universit\"{a}t Z\"{u}rich,
Winterthurerstrasse 190, CH-8057 Z\"urich, Switzerland}
\author{Yoichiro Hamaya}
\affiliation{Department of Crystalline Materials Science, Nagoya University,
Nagoya 464-8603, Japan}
\author{A.~Kaminski}
 \affiliation{Ames Laboratory and Department of Physics and Astronomy, Iowa
State University, Ames, IA~50011, USA}
%
%
\author{S.L.~Lee}
 \affiliation{School of Physics and Astronomy, University of St. Andrews, Fife,
KY16 9SS, UK}
\author{S.J.~Ray}
 \affiliation{School of Physics and Astronomy, University of St. Andrews, Fife,
KY16 9SS, UK}
\author{Tsunehiro Takeuchi}
\affiliation{Department of Crystalline Materials Science, Nagoya
University, Nagoya 464-8603, Japan}
\affiliation{EcoTopia Science
Institute, Nagoya University, Nagoya 464-8603, Japan}

\begin{abstract}
The superfluid density $\rho_s$ in underdoped ($T_c\simeq23$~K), optimally
doped ($T_c\simeq35$~K) and overdoped ($T_c\simeq29$~K) single crystalline
(BiPb)$_2$(SrLa)$_2$CuO$_{6+\delta}$ samples  was studied by means of muon-spin
rotation ($\mu$SR). By combining the $\mu$SR data with the results of ARPES
measurements on similar samples [Nature {\bf 457}, 296 (2009)] good
self-consistent agreement is obtained  between two techniques concerning the
temperature and the doping evolution of $\rho_s$.
\end{abstract}

\maketitle


The superfluid density $\rho_s$, being  proportional to the density of the
supercarriers, is one of the important characteristic of the superconducting
condensate.
Considering that superconductivity is characterized by the phase coherence of
electrons forming the pairs (two-particle process), it was previously believed that
techniques which probe the single-particle excitations of the condensate, such as photoemission
and single-electron tunneling, could not directly
provide the information on $\rho_s$.
It was quite unexpected, therefore, when angle-resolved photoemission (ARPES)
studies of the cuprate high-temperature superconductor (HTS)
Bi$_2$Sr$_2$CaCu$_2$O$_{8+\delta}$  revealed that a sharp peak, formed below
the superconducting transition temperature $T_c$ near the Brillouin zone
boundary, contains information not only on the pairing strength (the
superconducting energy  gap $\Delta$) but also on the phase coherence
\cite{Feng00,Ding02}. The peak intensity shows a clear resemblance to the
behavior exhibited by $\rho_s$ and scales linearly with $T_c$ in the underdoped
regime \cite{Feng00,Ding02}.

It should be mentioned that previous comparisons between  $\rho_s$ and the
coherence peak (CP) were made for CP measured near the Brillouin zone boundary
\cite{Feng00,Ding02}. The superfluid density, in its turn, is an ``angular
integrated'' quantity accumulating information over the full Fermi surface.
More importantly,  the carriers near the antinodes could be affected by the
pseudogap, which within ``two-gap'' scenario is supposed to be unrelated to the
pairing
 \cite{Chakravarty01,LeTacon06,Tanaka06,Kondo07,Hanaguri07,Guyard08,Yoshida09,Lee07, Kondo09}. This
could lead to an additional decrease of the CP intensity in the antinodal region
\cite{Kondo09}. Consequently the CP measurements over the full Brillouin zone
are needed in order to compare ARPES data with the superfluid density studied
independently in {\it e.g.}, muon-spin rotation ($\mu$SR) or microwave
experiments.

In this paper we report on the results of $\mu$SR studies of the superfluid
density and its comparison with the previously reported ARPES data
\cite{Kondo09} for underdoped ($T_c\simeq23$~K), optimally doped
($T_c\simeq35$~K) and overdoped ($T_c\simeq29$~K) single crystalline
(BiPb)$_2$(SrLa)$_2$CuO$_{6+\delta}$ samples. It was found that $\rho_s(T)$
could be well described with a superconducting gap of $d-$wave symmetry by
introducing the coherence quasiparticle weight function QW measured by means of
ARPES. The $T$ dependence of the superconducting energy gap $\Delta$ follows
the BCS prediction, and $\rho_s(T=0)$ scales with the CP intensity integrated
over the whole Fermi surface.


Details on the sample preparation of (BiPb)$_2$(SrLa)$_2$CuO$_{6+\delta}$
(Bi2201) single crystals can be found elsewhere \cite{Kondo09,Kondo04,Kondo05}.
The samples are labelled by their superconducting transition temperature $T_c$
with the prefix UD for underdoped, OP for optimally doped, and OD for overdoped
as: UD23K, OP35K, and OD29K.
\begin{figure}[htb]
\includegraphics[width=0.9\linewidth]{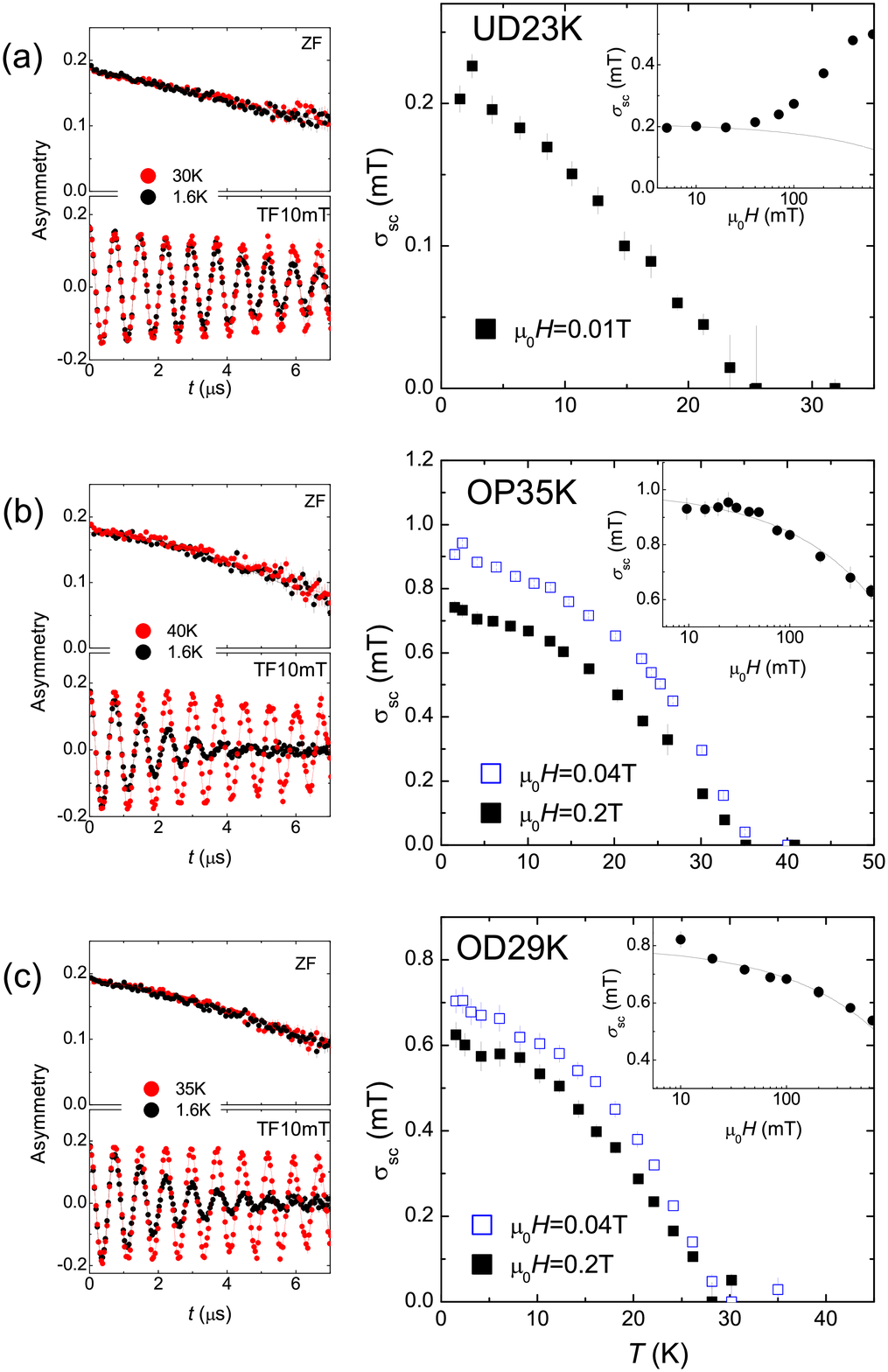}
%
\caption{(Color online) The representative ZF and TF-$\mu$SR
time-spectra measured above and below $T_c$ (left panels), and the
dependence of the superconducting part of square root of the
second moment $\sigma_{\rm
sc}\propto\lambda_{ab}^{-2}\propto\rho_s$ on temperature and
magnetic field (right panels) for UD23K (a), OP35K (b), and OD29K
(c).}
 \label{fig:muSR-raw-data}
\end{figure}
Zero-field (ZF) and transverse-field (TF) $\mu$SR experiments were carried out
at the $\pi$M3 beam line (Paul Scherrer Institute, Switzerland).
The samples were cooled from above $T_{\rm c}$ to 1.6~K at $H=0$ during
ZF-$\mu$SR experiments and in a series of fields ranging from 5~mT to 0.64~T in
TF-$\mu$SR experiments. In TF studies the magnetic field was applied parallel
to the $c$ axis and transverse to the muon-spin polarization. The typical
counting statistics were $\sim15-18$ million muon detections per data point.
The TF-$\mu$SR data for OP35K are partially published in
Refs.~\onlinecite{Khasanov08_Bi2201,Khasanov09_Bi2201}.

The representative ZF and TF-$\mu$SR time-spectra measured above
and below $T_c$ are presented in the left panels of
Figs.~\ref{fig:muSR-raw-data}~a, b, and c. In ZF all three samples
show a slow, temperature independent decay thus implying that the
$T$ dependent relaxation observed in TF  experiments at  $T<T_c$
should be attributed to the inhomogeneous field distribution in a
superconductor in the vortex state. The ZF data for OP35K and
OD29K are well described by the Gaussian decay function $A^{\rm
ZF}(t)=A_0\exp(-\sigma_G^2t^2/2)$ which is caused by the dipolar
field arising from the nuclear magnetic moments ($A_0$ is the
initial asymmetry and $\sigma_G$ is the Gaussian depolarization
rate).
The exponential character of the muon polarization decay in UD23K, described as
$A^{\rm ZF}(t)=A_0\exp(-\Lambda t)$, is an indication of a weak magnetism which
is probably caused by the closed proximity to the static magnetic order
\cite{Russo07}.

The TF-$\mu$SR data were analyzed by using a two-component Gaussian fit of the
$\mu$SR time-spectra allowing to describe the asymmetric local magnetic field
distribution $P(B)$ in a superconductor in the vortex state: $A^{\rm
TF}(t)=\sum_{i=1}^2 A_i \exp(-\sigma_i^2t^2/2) \cos(\gamma_{\mu}B_i t+\varphi)$
\cite{Khasanov08_Bi2201, Khasanov05_RbOsO,Khasanov06_LiPdB,Khasanov07_La214}.
Here $A_i$, $\sigma_i$, and $B_i$ are the initial asymmetry, relaxation rate,
and mean field of the $i-$th component, $\gamma_\mu = 2\pi\times135.5342$~MHz/T
is the muon gyromagnetic ratio, and $\varphi$ is the initial phase of the
muon-spin ensemble. The weak magnetism detected in ZF experiments for UD23K was
taken into account by multiplying the fitting function by $\exp(-\Lambda t)$.
The superconducting part of the square root of the second moment $\sigma_{\rm
sc}\propto\lambda_{ab}^{-2}\propto\rho_s$ ($\lambda_{ab}$ is the in-plane
magnetic penetration depth) was further obtained by subtracting the normal
state nuclear moment contribution ($\sigma_{\rm nm}$) from the measured second
moment of $P(B)$ ($\sigma^2$) as $\sigma_{\rm sc}=\sqrt{\sigma^2-\sigma_{\rm
nm}^2}$ \cite{Khasanov05_RbOsO,Khasanov06_LiPdB, Khasanov07_La214}.


The magnetic field dependence of $\sigma_{sc}$ at $T=1.6$~K for
the samples studied is shown in the corresponding insets of
Figs.~\ref{fig:muSR-raw-data}~a, b, and c. The decrease of
$\sigma_{sc}$ for OP35K and OD29K is a consequence of both, the
nonlinear and the nonlocal response of the superconductor
containing nodes in the energy gap to the increasing magnetic
field \cite{Khasanov09_Bi2201}. The solid lines for OD29K and
OP35K correspond to fits of the relation
$\sigma_{sc}(H)/\sigma_{sc}(H=0)=1-K\cdot\sqrt{H}$ to
$\sigma_{sc}(H)$ which takes into account the nonlinear correction
to $\rho_s$ for a superconductor with a $d-$wave energy gap
\cite{Vekhter99}. The analysis of $\sigma_{sc}(H)$ for UD23K
reveals, however, that only at very low fields ($\lesssim 30$~mT)
$\sigma_{sc}$ follows the tendency observed for OP doped and OD
Bi2201 samples. For higher fields $\sigma_{sc}$ {\it increases}
with increasing $H$. Such behavior is generally associated with
the field induced magnetism and is often observed in various
underdoped cuprate HTS (see {\it e.g.}
Refs.~\onlinecite{Khasanov07_NCCOC,Savici02}).

The observation of the field induced magnetism in UD23K for fields
exceeding 30~mT is quite unexpected, especially considering the
fact that in their recent study Russo {\it et al.} \cite{Russo07}
do not detect {\it any} kind of field induced effects in Bi2201
with $T_c\simeq27$~K up to $\mu_0H\simeq5$~T. We may suggest,
therefore, that the partial substitution of Bi by Pb, as made in
our samples, leads to enhancement of coupling between CuO$_2$
layers, thus causing Bi2201 to be more 3-dimensional. This could
affect both, the superconducting and the magnetic properties. As
for superconductivity, our previous studies point to a substantial
reduction of the anisotropy coefficient
$\gamma_\lambda=\lambda_c/\lambda_{ab}$ ($\lambda_c$ is the out-of
plane component of the magnetic penetration depth) as well as to
the shift of the vortex-lattice melting transition much closer to
$T_c$ \cite{Khasanov08_Bi2201}.  The magnetic properties could be
also affected due to increase of the interlayer magnetic exchange
coupling $J'$, which for the layered materials like cuprate HTS is
the primary quantity determining the N\'{e}el temperature
$T_N\propto J' \; \xi_{\rm 2D}^2$ ($\xi_{\rm 2D}$ is the magnetic
correlation lengths within the layer).

\begin{figure}[htb]
\includegraphics[width=0.9\linewidth]{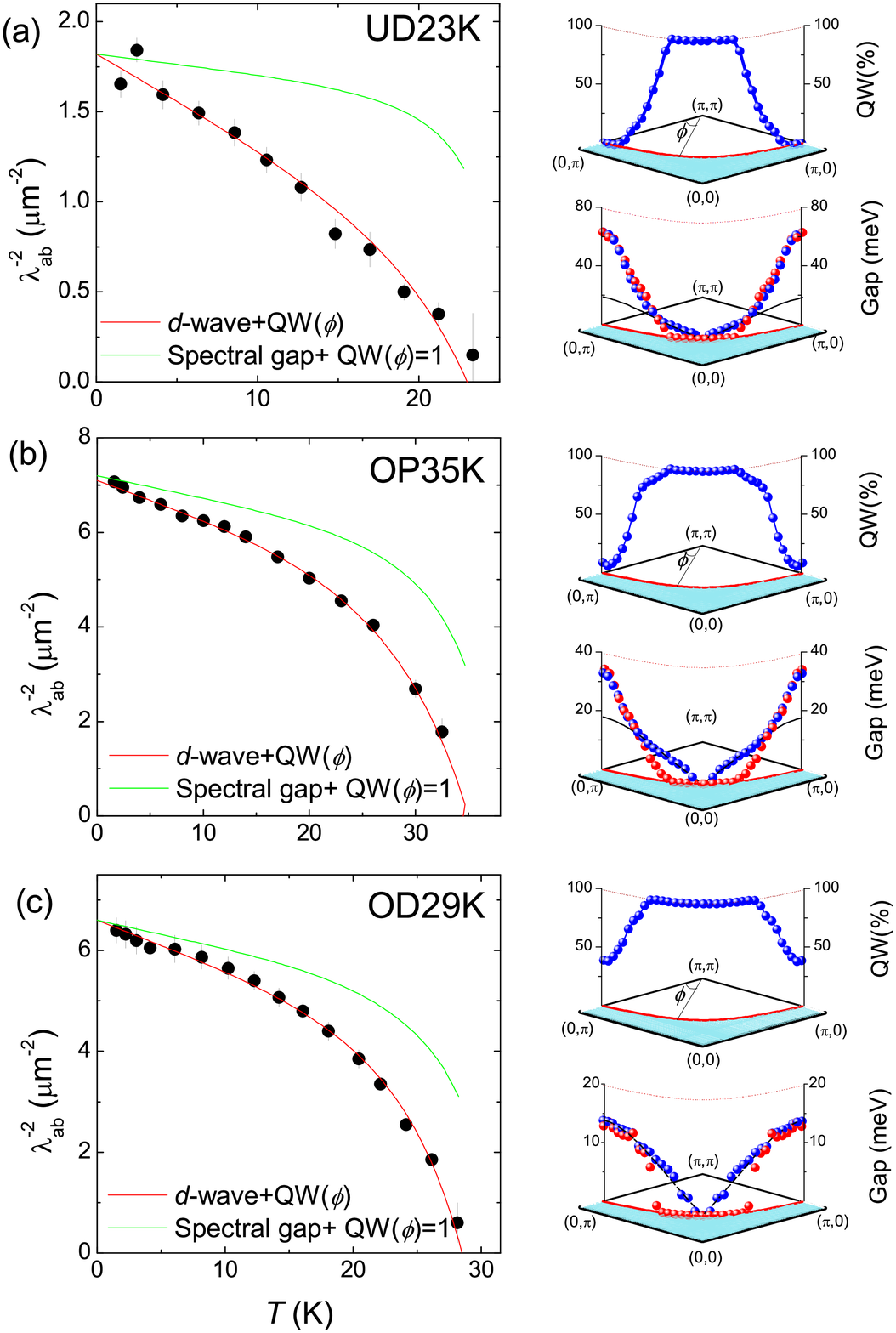}
\caption{(Color online) (a) The $\mu$SR and ARPES data for UD23K
sample. Left panel: $\lambda_{ab}^{-2}(T)\propto\rho_s(T)$. Lines
are the simulated $\lambda_{ab}^{-2}(T)$ curves obtained by
assuming that the full spectral gap measured by means of ARPES
leads to the pairing and QW$(\phi)=1$ (green line) and for the gap
of $d-$wave symmetry with QW$(\phi)$ measured by ARPES (red line).
Top right panel: angular dependence of the quasiparticle weight
function QW$(\phi)$ \cite{Kondo09}. $\phi=0$ and $\phi=\pi/4$
correspond to the antinodal and the nodal directions,
respectively. Bottom right panel: angular dependence of the
spectral gap above (41~K, red points) and below (11~K, blue
points) the transition temperature $T_c\simeq23$~K \cite{Kondo09}.
The solid line is the $\cos2\phi$ fit of $T=11$~K data in the
nodal region.
 (b) and (c) -- the same as in (a), but for OP35K and
OD29K, respectively.}
 \label{fig:lambda_vs_T}
\end{figure}

The temperature dependence of $\lambda_{ab}$ was obtained from the measured
$\sigma_{sc}(H=const,\; T)$'s shown in Fig.~\ref{fig:muSR-raw-data} by
following the procedure described in Ref.~\onlinecite{Khasanov09_Bi2201}. It
includes, first, the reconstruction of the effective penetration depth
$\lambda_{eff}(H,T)$ from $\sigma_{sc}(H,T)$ by using the relation
$\sigma_{sc}(H,T)=4.83\cdot10^4 [1 - H/H_{c2}(T)] [1 + 1.21(1 -
\sqrt{H/H_{c2}(T)})^3]\; \lambda_{eff}^{-2}$ \cite{Brandt03}. Here $H_{c2}$ is
the upper critical field with the zero-temperature values
$\mu_0H_{c2}(0)\simeq60$~T, $\simeq50$~T and $\simeq45$~T for UD23K, OP35K and
OD29K, respectively \cite{Wang03} and with the $T$ dependence following the
Werthamer-Helfand-Hohenberg prediction \cite{Werthamer66}. As a next step,
$\lambda_{ab}$ was reconstructed by decomposing $\lambda_{eff}(H,T)$  into the
field and the temperature dependent components as $\lambda_{\rm
eff}(H/H_{c2},T)=C(H/H_{c2})\;\lambda_{ab}(T)$.

The dependence of $\lambda_{ab}^{-2}\propto\rho_s$ on temperature is shown in
Fig.~\ref{fig:lambda_vs_T}. For OP35K and OD29K $\lambda_{ab}^{-2}(T)$ was
reconstructed from $\sigma_{sc}(H,T)$'s measured at $\mu_0H=0.04$, 0.1, 0.2,
0.4 and 0.64~T (see Ref.~\onlinecite{Khasanov09_Bi2201}), and $\mu_0H=0.04$ and
0.2~T, respectively. The field induced magnetism  does not allow us to make a
similar reconstruction for UD23K. We assumed, therefore,
$\lambda_{ab}(T)\simeq\lambda_{eff}(0.01$~T$, T)$. As shown in
Ref.~\onlinecite{Khasanov09_Bi2201} the relation
$\lambda_{eff}(H,T)\simeq\lambda_{ab}(T)$ is still valid in the limit of low
magnetic fields $H\lesssim 10^{-3} H_{c2}(0)$.
\begin{figure}[htb]
\includegraphics[width=1.0\linewidth]{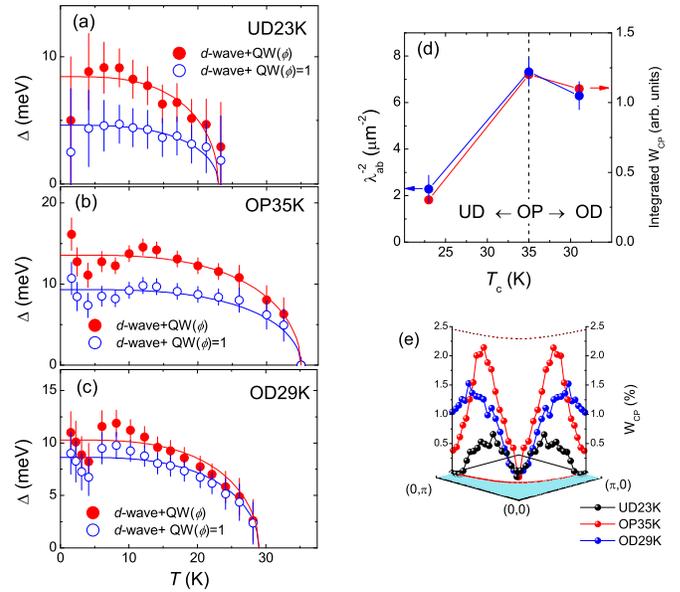}
\caption{(Color online) Temperature dependence of the
superconducting gap for UD23K (a), OP35K (b), and OD29K (c). The
open and closed symbols correspond to QW$(\phi)=1$ and QW$(\phi)$
measured by means of ARPES \cite{Kondo09}. The solid lines are BCS
fits with parameters summarized in Table~\ref{Table:Gap-results}.
(d) Dependence of $\lambda_{ab}^{-2}(T=0)$ and the coherence peak
intensity $W_{\rm CP}$ integrated over the whole Fermi surface on
$T_c$.
(e) Angular dependence of $W_{\rm CP}$ measured by ARPES \cite{Kondo09}.}
 \label{fig:gaps}
\end{figure}

The resulting $\lambda_{ab}^{-2}(T)$'s  were analyzed within the same scheme as
described in Refs.~\onlinecite{Khasanov08_Bi2201,Khasanov09_Bi2201}:
\begin{equation}
\frac{\lambda_{ab}^{-2}(T)}{\lambda_{ab}^{-2}(0)}=  1+ \frac{2}{S_{\rm QW}}
\int_{0}^{\pi/4}\int_{\Delta(T,\phi)}^{\infty}\left(\frac{\partial f}{\partial
E}\right)\frac{{\rm QW}(\phi)\  E\ dEd\phi}{\sqrt{E^2-\Delta(T,\phi)^2}}\ .
 \label{eq:lambda-d}
\end{equation}
Here $f=[1+\exp(E/k_BT)]^{-1}$ is the Fermi function, $\phi$ is the angle along
the Fermi surface (see the top right panels in Figs.~\ref{fig:lambda_vs_T}~a,
b, and c), $\Delta(T,\phi)$ denotes the superconducting (pairing) gap depending
on $T$ and $\phi$, QW($\phi$) accounts for the relative weight of the
quasiparticles condensed into the Cooper pairs, and $S_{\rm
QW}=\int_0^{\pi/4}{\rm QW}(\phi)d\phi$. Note that we do not consider here the
presence of the second $s-$wave gap
\cite{Khasanov07_La214,Khasanov07_Y123,Khasanov08_Y124} since the $d-$wave
contribution to $\lambda_{ab}^{-2}(T)$ in Bi2201 seems to be predominant.

As a first step we checked if the full spectral gap measured in ARPES
experiments (lower right panels in Figs.~\ref{fig:lambda_vs_T}~a, b, and c)
could be a pairing gap. The solid green lines on the left panels of
Figs.~\ref{fig:lambda_vs_T}~a, b, and c correspond to the theoretical curves
obtained by assuming that  the states over the whole Fermi surface are equally
available for condensation [QW$(\phi)=1$], and that the gap in the antinodal
region is $T$-independent and it follows the BCS prediction close to the nodes
\cite{Lee07}. The poor agreement between the experiment and the theory suggests
that the gap near the antinodes could not be related to the pairing. In
particular, the step-like jump of $\rho_s$ at $T=T_c$ is due to the fact that
the antinodal gap is not closed when the temperature passes through $T_c$.

We suggest, therefore, that for all levels of doping the
superconducting (pairing) gap has a $d-$wave symmetry:
$\Delta(T,\phi)=\Delta(T)\;\cos2\phi$. This is indeed true for
OD29K (Fig.~\ref{fig:lambda_vs_T} and Ref.~\onlinecite{Kondo09}),
as well as for various OD and OP hole-doped cuprate HTS studied by
means of ARPES (see {\it e.g.} Ref.~\onlinecite{Damascelli03} and
references therein). The predominantly $d-$wave symmetry of the
order parameter in UD HTS was also confirmed in tricrystal
experiments \cite{Tsuei04}. This, together with known QW$(\phi)$
and $\lambda_{ab}^{-2}(T)$ allows the use of
Eq.~(\ref{eq:lambda-d}) to reconstruct $\Delta(T)$. The results of
such reconstruction for two possible scenarios are shown in
Fig.~\ref{fig:gaps}: the first with QW$(\phi)$ obtained by means
of ARPES (see Fig.~\ref{fig:lambda_vs_T} and
Ref.~\onlinecite{Kondo09}), and the second with QW$(\phi)=1$,
indicated by the closed and open symbols respectively. Fits of the
BCS model to $\Delta(T)$ are represented by solid lines. Both sets
of $\Delta(T)$ data follow the BCS temperature dependence in
agreement with the results
of ARPES studies for the spectral gap in the nodal region \cite{Lee07}.

\begin{table}[htb]
\caption[~]{\label{Table:Gap-results} Zero-temperature values of the
superconducting gap $\Delta(0)$ and the ratio $2\Delta(0)/k_BT_c$ as obtained
from the fit of the BCS model to $\Delta(T)$ represented in
Figs.~\ref{fig:gaps}~a, b, and c. The ''$d-$wave+ QW$(\phi)=1$`` and
''$d-$wave+QW$(\phi)$`` refer to the case of QW$(\phi)=1$ and QW$(\phi)$
measured by means of ARPES \cite{Kondo09}, respectively (see text for
details).}
\begin{center}
\begin{tabular}{lccc|ccc}\\
 \hline
 \hline
& \multicolumn{2}{c}{$d-$wave+ QW$(\phi)=1$}&&& \multicolumn{2}{c}{$d-$wave+QW$(\phi)$}\\
&$\Delta(0)$&$\frac{2\Delta(0)}{k_BT_c}$&&&$\Delta(0)$&$\frac{2\Delta(0)}{k_BT_c}$\\
&(meV)&&&&(meV)&\\
\hline
UD23K&4.6(2)&4.6(2)&&&8.4(3) &8.5(3)\\
OP35K&9.3(2)&6.2(1)&&&13.6(3)&9.0(2)\\
OD29K&8.6(2)&6.9(2)&&&10.3(3)&8.2(2)\\
 \hline \hline \\

\end{tabular}
   \end{center}
\end{table}

In order to distinguish between the two above mentioned scenarios we note
that:
(i) Accounting for the quasiparticle weight as measured by ARPES causes a
systematic shift of $\Delta(T)$ to higher values. This leads to better
agreement with the gap obtained from the $\cos2\phi$ fit to the ARPES data in
the nodal region \cite{Kondo09}.
(ii) The ratio $2\Delta/k_BT_c$ increases with doping from $\simeq4.6$ for
UD23K to $6.9$ for OD29K in a case when QW$(\phi)=1$ while it stays almost
constant ($\sim 8.5$) for QW$(\phi)$ obtained by means of ARPES, see
Table~\ref{Table:Gap-results}. Note that the independence of  $2\Delta/k_BT_c$
ratio on doping is well confirmed experimentally for various HTS families (see
e.g. Refs.~\onlinecite{Khasanov08_OIE_on_Gap,Guyard08_2} and references
therein).
(iii) The intensity of the coherence peak $W_{\rm CP}$ (Fig.~\ref{fig:gaps}~e)
integrated over the whole Fermi surface scales with the zero-temperature
superfluid density (Fig.~\ref{fig:gaps}~d), thus pointing to the direct
relation of $W_{\rm CP}$ to the ''local`` superfluid density
\cite{Feng00,Ding02,Kondo09}. By approaching the antinodal point, $W_{\rm CP}$
decreases thus requiring the corresponding decrease in the local density of the
supercarriers. The strongest effect (decrease of $W_{\rm CP}$ down to 0) is
observed for UD23K while for OD29K the weight of CP in the nodal region still
remains substantial.
All these arguments taken together support the scenario according to which the
gap (pseudogap) near the antinodes makes a part of the states unavailable for
the superconducting condensation and leads, therefore, to a reduced density of
the supercarriers in the antinodal region.

To summarize, the temperature dependence of the superfluid density $\rho_s$ was
studied in underdoped ($T_c\simeq23$~K), optimally doped ($T_c\simeq35$~K) and
overdoped ($T_c\simeq29$~K) single-crystalline Bi2201 samples by means of
muon-spin rotation. By comparing the measured $\rho_s(T)$ with that calculated
theoretically based on the results of ARPES  \cite{Kondo09} we found that the
superconducting gap in Bi2201 at all levels of doping has $d-$wave symmetry and
that $\Delta(T)$ follows reasonably well the BCS prediction. It was also shown
that $\rho_s(T)$ is inconsistent with the case when the carriers over the whole
Fermi surface are equally available for condensation thus suggesting that some
parts of the Fermi surface do not develop the superconducting coherence.


This work was performed at the Swiss Muon Source (Paul Scherrer Institute,
Switzerland). Work at the Ames Laboratory was supported by the Department of
Energy - Basic Energy Sciences under Contract No. DE-AC02-07CH11358. The
financial support of the Swiss National Foundation (SNF) is gratefully
acknowledged.

\end{document}